\documentclass[11pt, letterpaper]{article} 

\usepackage[includeheadfoot,
            marginratio={1:1,2:3}, 
            width=440pt, 
            height=688pt,]{geometry}

\usepackage{amsmath}
\usepackage{amsfonts}
\usepackage{amssymb}
\usepackage{bbm,bm}
\usepackage{mathtools}
\usepackage{slashed}
\usepackage{mathalfa}

\usepackage{graphicx,caption,subfigure}
\usepackage{tikz} 
\usepackage{tikz-cd}
\usetikzlibrary{decorations.pathreplacing,calc,tikzmark}
\usepackage{booktabs}
\usepackage{adjustbox}
\usepackage[utf8]{inputenc}
\usepackage[splitrule]{footmisc} 
 
\usepackage{placeins}
\usepackage{empheq}
\usepackage{paralist}
\usepackage{cite}
\usepackage[normalem]{ulem}
\usepackage{soul}

\usepackage[colorlinks=false,urlbordercolor=red
]{hyperref}
\usepackage{xcolor}
\usepackage{tensor}

\newcommand{\nc}{\newcommand}
\nc{\lb}{\llbracket}
\nc{\rb}{\rrbracket}
\nc{\gl}{\llbracket}
\nc{\gr}{\rrbracket}
\nc{\del}{\partial}
\nc{\eq}[1]{\begin{equation}
                     \begin{split} #1 \end{split}
                     \end{equation}
}
\nc{\ov}{\overline}
\nc{\fa}{\hat}
\nc{\fb}{\MakeUppercase}
\nc{\fc}{\tilde}
\nc{\myhash}{\raisebox{\depth}{\#}}

\allowdisplaybreaks[2]
\numberwithin{equation}{section}
\DeclareUnicodeCharacter{2212}{-}

\begin{document}


\vspace*{-1.5cm}
\begin{flushright}
  {\small
  MPP-2026-80
  }
\end{flushright}

\vspace{1.0cm}
\begin{center}
  {\huge  Towards the Realization of the  Dark Dimension \\[0.4cm] Scenario  in Ho\v{r}ava-Witten Theory 
    } 
\vspace{0.4cm}

\end{center}

\vspace{0.25cm}
\begin{center}
{
\Large Ralph Blumenhagen and Antonia Paraskevopoulou

}
\end{center}

\vspace{0.0cm}
\begin{center} 
  \emph{ 
Max-Planck-Institut f\"ur Physik (Werner-Heisenberg-Institut), \\ 
Boltzmannstra\ss e  8,  85748 Garching, Germany } 
\\[0.1cm] 
\vspace{0.25cm} 
\vspace{0.3cm}
\end{center} 
\vspace{0.5cm}

\begin{abstract}
  It has been suggested that Ho\v{r}ava-Witten theory could provide
  a concrete realization of the Dark Dimension Scenario.
  In this context, the observable Standard Model sector is naturally localized
  in the micron-sized  large dimension, which is the 
  interval in the eleventh direction. Considering Calabi-Yau
  manifolds supporting generic vector bundles  including also abelian factors, we point out  that
    symmetric tadpole cancellation on the $E_8$ walls  has the
    potential to  ameliorate some
  of the issues of such a realization, including too fast proton decay.
   By taking not only the hierarchically small value 
of the dark energy but also the size of the Standard Model gauge couplings into account,
  one is driven to a  special   infinite distance limit, which is the
  Ho\v{r}ava-Witten analogue of a limit recently at the focus of  the M-theoretic  Emergence Proposal.
  Extrapolating results obtained for BPS-saturated amplitudes, we
  speculate about  the possibility of obtaining the moduli dependence of
  the scalar potential, the gauge couplings and the Planck scale
  by  simple one-loop Schwinger
  integrals over  towers of states. 
 \end{abstract}

\thispagestyle{empty}
\clearpage

\setcounter{tocdepth}{2}


\section{Introduction}

The swampland program has
revealed that string theory can be more predictive than commonly
thought (see
\cite{Palti:2019pca,vanBeest:2021lhn,Grana:2021zvf,Agmon:2022thq} for reviews). This is due
to a new concept of naturalness claiming that only those effective quantum field theories which admit a consistent ultraviolet (UV)
completion are natural. This approach focuses on the low energy imprints of quantum gravity, which are by definition manifestations of ultraviolet/infrared  (UV/IR) mixing.

A very prominent example of such an approach is the Dark Dimension
Scenario \cite{Montero:2022prj} (see \cite{Vafa:2024fpx,Anchordoqui:2024ajk} for reviews).
Its starting point is the assumption that our universe is located in
a  region of field space where  the generalization of the Anti-de Sitter (AdS) 
distance conjecture \cite{Lust:2019zwm} also applies to
de Sitter (dS) space. This means that 
for the Dark Energy density  $\Lambda_{\rm DE}$ approaching zero, a tower
of states becomes light, obeying the specific scaling behavior
\eq{  \label{dSdistance}
     m\sim \Lambda_{\rm DE}^\alpha\,,}
with the exponent lying in the range $1/4\le \alpha\le 1/2$.
Applying this to our universe with its observed tiny cosmological
constant $\Lambda_{\rm DE} = 10^{−122} M_{\rm pl}^4$
and taking bounds from deviations of Newton’s gravitational force law into
account, revealed that only the lower bound  $\alpha = 1/4$ is consistent.
This bound arose from taking one-loop corrections from integrating
out the tower of light states into account.
Using \eqref{dSdistance} one finds that the mass scale of this tower is
$m\sim 1\,$meV. The emergent string conjecture \cite{Lee:2019wij} claims
that these lightest modes are either (dual to) Kaluza-Klein (KK)
modes or excitations of a weakly coupled critical string.
Since the plethora of  string excitations has not been observed at
such a scale, the only remaining possibility are KK modes.
Astrophysical constraints from the cooling/heating of neutron stars,
similarly to the original Large Dimension Scenario \cite{Arkani-Hamed:1998sfv},
 led to a model of a single large
extra dimension, dubbed the Dark Dimension, and a corresponding tower
of KK modes. However, see also \cite{Anchordoqui:2025nmb,Hardy:2025ajb} for considerations of two large extra dimensions.

An appealing aspect of this scenario is that the KK modes of the graviton
can be natural dark matter candidates \cite{Gonzalo:2022jac}.
Alternatively, also primordial black holes could play
a role  \cite{Anchordoqui:2022txe}.
In addition, the recent results of the DESI collaboration \cite{DESI:2025zgx,DESI:2025fii} indicate
that the dark energy might be decreasing with time, something
that can be seen as a prediction of this scenario where
the size of the single large extra dimension rolls down
its potential \cite{Bedroya:2025fwh}.
Since one does not observe light KK modes of Standard Model
fields, it is clear that the  Standard Model  must be realized
on a brane that is localized in the large extra dimension.

Despite its implementation of swampland conjectures, which are mainly motivated by string theory,
it is challenging to find a concrete string theoretic
realization of this scenario with a single unstabilized
direction (see \cite{Blumenhagen:2022zzw,Schwarz:2024tet,Dudas:2025yqm} for recent attempts).
For instance, when trying  to construct an intersecting $D$-brane model 
in 10D perturbative string theory, one faces the problem
of singling out the large extra dimension and stabilizing
all the remaining moduli of the orthogonal   five dimensional manifold  via e.g.~fluxes and instantons.

As proposed in \cite{Schwarz:2024tet}, from this
perspective eleven dimensional M-theory or Ho\v{r}ava-Witten (HW)
theory \cite{Horava:1995qa,Horava:1996ma} appears
as a more natural candidate, in particular when
considering  it as the strong coupling limit of type IIA theory,
or the $E_8\times E_8$ heterotic string, respectively.
Here, one compactifies M-theory on a seven dimensional space of the type
\eq{
  X\times S^1\,, \qquad\qquad\quad    X\times {\frac{S^1}{\mathbb Z_2}}\,,
  }
  where the six-dimensional compact space  $X$ can be e.g.~a 
  Calabi-Yau manifold, which are well understood and leads in the HW case
  to theories with $N=1$ supersymmetry in four dimensions.
  Now, if one can really keep  this product structure and not solely
  a fibration or a warped product, then the singled out eleventh direction
  provides a natural candidate for the large extra dimension. 
  Moreover, the HW theory also provides two codimension-one
  end-of-the-world branes where one can imagine that
  the Standard Model lives.

This paper is organized as follows.
In section \ref{sec2}, we briefly review HW theory and address
some of the concerns raised about such a construction, like the backreaction of the $E_8$ walls
on the eleventh direction,  related to a strong coupling singularity
on one of the two walls,  or the problem of  too fast proton decay.
Our analysis suggests that, unless higher order obstructions arise,
implementing what we will call symmetric tadpole
cancellation on the $E_8$ walls  (for earlier studies see 
\cite{Lukas:1999de}) and allowing general vector bundles
involving also abelian factors in their structure group,
can ameliorate these problems. Concerning proton decay, in an
appendix we provide a concrete heterotic quiver model
as a proof of principle that heterotic/HW
MSSM models with  approximate global lepton and baryon number
symmetries can exist.
This analysis implies that HW theory remains a promising arena for  a string theory realization
of the Dark Dimension Scenario.

Due to the large eleventh direction, one is in a decompactification limit of 4D quantum gravity.
Taking also the size of the Standard Model gauge couplings
into account, we will see  that this limit
is precisely of the type that was recently considered in the context of the
Emergence Proposal
\cite{Heidenreich:2017sim,Grimm:2018ohb,Heidenreich:2018kpg}
(see e.g.~also \cite{Marchesano:2022axe,Castellano:2022bvr,Blumenhagen:2023yws,Hattab:2023moj,Hattab:2024thi,Hattab:2024ssg}), where in particular the size of the Calabi-Yau manifold is of order
one in M-theory units.
In fact, it is the limit that was called the M-theory limit in
\cite{Blumenhagen:2023tev,Blumenhagen:2023xmk,Blumenhagen:2024ydy,Artime:2025egu}
but also appeared before in the classification of large
distance limits \cite{Corvilain:2018lgw}, called type IV$_d$ .
In those papers it was shown in detail that certain BPS-saturated effective couplings
were given {\it exactly} by single Schwinger integrals over
the light towers of BPS states (see also \cite{Blumenhagen:2024lmo}
for a review). From the weakly coupled
string point of view, this included tree-level, loop and
instanton corrections.

Extrapolating this result to the Dark Dimension limit of  HW theory,
in section \ref{sec:emergence} we take a few steps
towards extracting information about the form
of certain quantities in this genuine M-theoretic regime, namely the
scalar potential, the gauge couplings  and  the 4D Planck mass.
Since we do lack a full description of quantum M-theory,
these results are, at the moment, speculative but may provide a
valuable first glimpse
into this regime.

\section{The Dark Dimension Scenario in Ho\v{r}ava-Witten Theory}
\label{sec2}
In this section, we recall  a few salient points which will be relevant for the realization
of the Dark Dimension Scenario in HW Theory.

\subsection{Aspects of Ho\v{r}ava-Witten Theory}

HW theory \cite{Horava:1995qa,Horava:1996ma} is the strong coupling limit of the $E_8\times E_8$
heterotic string. It can be described as 11D M-theory
compactified on a circle $S^1$ and then a quotient by
a $\mathbb Z_2$ symmetry. This acts like
$x_{11}\to -x_{11}$ on the eleventh direction
and also flips the sign of the three-form gauge field,
$C_3\to -C_3$.
Hence, one has M-theory on an interval $I_1=S^1/\mathbb Z_2$ of size
$2\pi R_{11}=\pi R^{\rm HW}_{11}$,
where anomaly cancellation on the fixed points 
of this orbifold requires the presence  of an $E_8$ gauge group
at each of the two ends of the world\footnote{In the following we will use
  $R_{11}$ so that the relations take a form familiar from M-theory on a circle.}. Similarly to the strong
coupling limit of type IIA, the heterotic string coupling is related
to the size of the interval via
\eq{  g_s=(M_* R_{11})^{{3/2}}\,,                   }
 where $M_*$ denotes the eleven dimensional Planck scale\footnote{We
   use the standard relation to the coefficient of the 11D
   Einstein-Hilbert action $2\kappa_{11}^2={\frac{1}{2\pi}}\big({\frac{2\pi}{M_*}}\big)^9$.}.
 The latter is also related to the heterotic string scale $M_s$ via
 \eq{
   \label{hannover96}
                        M_s^2=M_*^3 R_{11}\,.
                      }
                      
The heterotic string as well as HW theory have been suggested
to be good candidates for string theory realizations
of supersymmetric Grand Unified Theories (GUTs) in four dimensions.
For this purpose, one compactifies the theory on a Calabi-Yau
threefold leading to $N=1$ supersymmetry in four dimensions.
This is also the framework that we will consider in this paper.
As usual in string model building, it can be considered
as a better controlled intermediate step from which  the final
quasi dS minimum should arise via supersymmetry breaking effects
of fluxes and instantons.
For instance, the common GUT gauge groups $SU(5)$ and $SO(10)$
are both subgroups of $E_8$, so
that one can realize them by turning on a non-trivial $SU(5)$ or 
$SU(4)$ gauge bundle on one of the two $E_8$ factors.
The breaking to the Standard Model gauge group can
then be achieved by turning on appropriate discrete Wilson lines.

In general 
one could turn on non-trivial background gauge bundles
in both $E_8$ factors.  
In this case, the tadpole cancellation condition is that
the combination
\eq{\label{tadpole1}
            dH_3\sim    {\rm Tr}(F_1\wedge F_1)+{\rm Tr}(F_2\wedge F_2) -{\rm Tr}(R\wedge R)
}
is trivial in cohomology, so that
\eq{\label{tadpole2}
         {\rm ch}_2(V_1)+ {\rm ch}_2(V_2)+c_2(T_X)=0\,.
}
Here, $F_1$ and $F_2$ are the two gauge
backgrounds of the vector bundles $V_1$ and $V_2$
on the two end-of-the-world domain walls, $R$  denotes  the
curvature two-form and we assumed  the absence of space-time filling 5-branes.

Moreover, one could not only allow for  gauge bundles with trivial
first Chern class, like the  aforementioned $SU(5)$ or 
$SU(4)$ bundles, but also  
consider more general constructions
of a Standard Model sector, like the ones involving
turning on non-trivial line bundles in both $E_8$
 factors
\cite{Distler:1987ee,Lukas:1999nh,Blumenhagen:2005ga,Blumenhagen:2006ux,Anderson:2011ns}.
In this way, the model
building possibilities increase dramatically. In fact,
they become similarly rich to those for intersecting $D$-brane models.
For instance, one could directly start with a bundle with structure
group $SU(5)\times U(1)$ so that the commutant in $E_8$ is
$SU(3)\times SU(2)\times U(1)$.

Of general concern is the following issue in HW theory. In \cite{Witten:1996mz},
 it was shown that turning on general non-trivial  gauge bundles $F_i$
 in  the two $E_8$ factors induces  $M5$-brane charges that
 act as source terms in the Bianchi identify for the four-form
 flux and leads to a warped compactification.
Then the size of the Calabi-Yau $X$ depends on the position $x_{11}$
so that in the linear approximation\footnote{In \cite{Cvetic:2024wsj} it was pointed out that taking 
  instanton corrections into account, the singularity might be
  resolved so that one can continue $R_{11}$ beyond this boundary
  value (see also \cite{Curio:2000dw} for earlier studies going beyond
  the linear approximation). However, it is not clear how this new regime precisely
  looks like and if e.g.~it still has a large extra dimension.} one encounters  a critical  distance
\eq{
  \label{wittenrelation}
  {\frac{1}{M_* R_{11}^{\rm crit}}}=  {\frac{\lambda}{(M_* R_{\rm
        CY})^{6}}} \bigg\vert{\frac{M_*^2}{8\pi^2}}\int_X J\wedge
  \left({\rm Tr}(F_i\wedge F_i)- {\frac{1}{2}}{\rm Tr}(R\wedge R)\right)\bigg\vert\,,
}
 around which (quasi-classically) the size of the  Calabi-Yau manifold
 becomes negative. 
Here the numerical factor is of order $\lambda=1/(8\pi^2)\sim 10^{-2}$
and $J$ denotes the K\"ahler form of the Calabi-Yau. 
As discussed in \cite{Reig:2025dpz} and as we will see in the next section,
for the Dark Dimension scenario with a wanted micron size
eleventh direction, this linear  backreaction effect is
lethal. The appearance of this singularity  is consistent with the recent analysis of generic
obstructions of the infinite distance limits in the moduli space of   4D $N=1$ supersymmetric
compactifications \cite{Kaufmann:2026fli,Kaufmann:2026mha,Kaufmann:2026tsy}.

The only obvious  way to avoid this issue from the very beginning is by turning
on   gauge bundles in the two $E_8$ factors satisfying 
\eq{
  \label{localtadpole}
       {\rm Tr}(F_i\wedge F_i)- {\frac{1}{2}}{\rm Tr}(R\wedge R)
       =0\,,\qquad i=1,2
}     
in cohomology. Then, taking $dJ=0$ into account, the right hand
side of \eqref{wittenrelation} is vanishing.
Note that this does not imply that the two gauge bundles are the same,
not even that they have the same structure group.
Nevertheless, we call this simply  ``symmetric tadpole cancellation'',
understanding that it is only meant in the above topological sense.
Apparently, the standard embedding is not
an admissible  solution but
concrete realizations of such models (called ``symmetric vacua") have been investigated
in \cite{Lukas:1999de}.

Even though  the absence of warping in the $x_{11}$ direction and the
resulting linear obstruction
is certainly
a nice feature, it is not guaranteed at this level whether 
further  quantum obstructions to
the large $R_{11}$ regime  can occur, e.g.~like  those studied in \cite{Kaufmann:2026fli,Kaufmann:2026mha,Kaufmann:2026tsy}.
It is not the aim of this paper to fully resolve this issue but we
trust the intuition that symmetric tadpole cancellation implies
that the backreaction is only along the two $E_8$ walls and does not
occur along the $x_{11}$ direction. Hence, we are really
having a trivially fibered  product space $X\times I_1$ so that
the large  $R_{11}$ regime is part of the HW moduli space.

\subsection{Aspects of the weakly coupled heterotic string}

Even though eventually the eleventh direction is large
and one needs to work in the strong coupling regime of the
heterotic string, it is useful to recall some results
for the weakly coupled heterotic string, which
will become relevant in the next section.
In particular, the results depending  on topological
quantities like the massless chiral spectrum
or the Green-Schwarz mechanism, i.e.~anomaly cancellation,
are expected to also hold in the strong coupling limit \cite{Lalak:1998jg,Lukas:1999nh}.

By dimensionally reducing the 10D action of the
heterotic string on a Calabi-Yau threefold, one gets the tree-level
(reduced)  Planck-scale in 4D
\eq{
  M_{\rm pl}^2 = {\frac{1}{\pi}}\frac{M_s^2}{g_s^2} {\cal V}_6\,,
  }
  where ${\cal V}_6=V_6/(2\pi \sqrt{\alpha'})^6$ denotes the volume of the Calabi-Yau in string units.

  The gauge coupling, respectively the holomorphic
gauge kinetic function, of a non-abelian gauge factor in the i-th
$E_8$ factor satisfies a non-renormalization theorem and enjoys a schematic  expansion
\eq{
  \label{gaugekinf}
       4\pi f^{(i)}_{\rm YM} = S + \gamma \, \frac{M_s^2}{8\pi^2}\int_X
       {J}\wedge \left({\rm Tr}(F_i\wedge F_i)- {\frac{1}{2}}{\rm
           Tr}(R\wedge R)\right) + \mathcal{O}\big(e^{- 2\pi T_k}\big) + \mathcal{O}\big(e^{-2\pi S}\big) \,,
}
where we defined the  complex moduli fields
\eq{
            S= \frac{{\cal V}_6}{g_s^2} + i \int_X B_6\,,\qquad\quad
            T_k= {\frac{1}{(2\pi \sqrt{\alpha'})^2}} \int_{\Gamma_k} { J}
            + i \int_{\Gamma_k} B_2\,,
          }
for a basis of homological 2-cycles   $\Gamma_k\in H_2(X,\mathbb Z)$.
The first term in \eqref{gaugekinf} is at tree-level resulting from the dimensional
reduction of the gauge kinetic term in 10D, the second and third term
are a
one-loop correction \cite{Ibanez:1986xy} which contains  also world-sheet instanton
contributions
and the last term indicates potential $N\!S5$-brane instanton
corrections.

Let us remind the reader that due to the tadpole cancellation condition
\eqref{tadpole1}, for non-symmetric models the second term in
\eqref{gaugekinf} is negative
on one of two $E_8$ walls \cite{Banks:1996ss}.
This implies that  at the critical radius
\eqref{wittenrelation} its gauge coupling diverges, unless
the instanton corrections cure this strong coupling  singularity as in \cite{Cvetic:2024wsj}.
Again, this issue is avoided upon imposing  symmetric  tadpole
cancellation \eqref{localtadpole}.

Allowing also abelian gauge bundles $L_a$  will result in abelian
gauge groups in 4D whose one-loop corrections will contain
additional terms \cite{Blumenhagen:2005ga}
\eq{
  \label{gaugeu1}
              \Delta {\rm Re}(f^{(i)}_{ab}) \sim \int_X  J\wedge
              c_1(L_a)\wedge c_1(L_b) \,,
 }
i.e.~there will also be non-diagonal couplings.
Note that these contributions             
are  generically non-vanishing even for  symmetric vacua,
but that their  signs are a priori not fixed.

As discussed in \cite{Blumenhagen:2005ga,Blumenhagen:2006ux},
relatedly one also
has to consider the Green-Schwarz mechanism, by which
abelian gauge factors can gain a mass and  
a non-vanishing
Fayet-Iliopoulos (FI)  term is generated for the corresponding 4D abelian $U(1)$ gauge group
\eq{
  \label{FIterm}
  \xi^{(i)}_a\sim \frac{M_s^4}{g_s^2} \int_X J\wedge J\wedge c_1(L_a) -
              \int_X c_1(L_a)\wedge \left( {\rm Tr}(F_i\wedge F_i)- {\frac{1}{2}}{\rm Tr}(R\wedge R)\right)+\ldots\,.
           }
Again the one-loop correction vanishes upon imposing  symmetric tadpole
cancellation.

\subsection{Realizing the Dark Dimension Scenario}

Even though the Dark Dimension Scenario provides an appealing and even predictive
framework for connecting cosmology with recent results
from the swampland program of quantum gravity,
it is not straightforward to come up with a convincing
top-down construction within a UV complete theory like string theory to realize it.
Recall that due to the large direction, in the Dark Dimension scenario the quantum
gravity cutoff,
the so-called species scale \cite{Veneziano:2001ah,Dvali:2007hz},  is  equal to the 5D Planck scale and of the order 
\eq{
  \label{werder}
  \Lambda_{\rm sp}\sim M_{\rm pl}^{2/3} m_{\rm KK}^{1/3}\sim
  10^9\,{\rm GeV}\,.
}

The motivation for considering HW theory is that it naturally offers a candidate for the single large extra dimension, namely the eleventh direction. Given that one requires this to be large, one is at the strong coupling limit of heterotic string theory.
 Our world with the  Standard Model of particle physics  should then
appear by choosing appropriate bundles on the end-of-the-world
$E_8$ walls. These are localized in the eleventh direction
and thus by construction do not lead to
fatal light KK modes of the Standard Model fields.

Now, we go beyond cosmology and implement as 
an additional observational input  that the gauge couplings
of the Standard Model are of the order $\alpha_{\rm SM}\sim 0.1$.
By dimensional reduction of the 10D Yang-Mills action to 4D,
one gets
\eq{
  \label{gaugescale4D}
                {\frac{1}{\alpha_{\rm SM}}}= {\frac{{\cal V}_6}{g_s^2}} \,,
}
which, of course,  is the tree-level term in \eqref{gaugekinf}.
Assuming an isotropic Calabi-Yau with $V_6=(2\pi R_{\rm CY})^6$,  we can express
the gauge coupling in M-theory units as
\eq{
                {\frac{1}{\alpha_{\rm SM}}}= (M_* R_{\rm CY})^6\,,
}
i.e.~as expected it does not depend on the large eleventh direction.
This constrains the Calabi-Yau radius to be of order one in 11D
Planck units with $r_{\rm CY}:=M_* R_{\rm CY}\sim 1.5$.
This is not parametrically in the large volume regime
where one  can safely trust an 11D  supergravity approximation
and it also means that the cutoff, i.e.~the 5D Planck scale, is of the order
of the 11D Planck scale.

These relations will  have  direct consequences for the whole
construction.  Dimensional reduction of the 11D supergravity action
leads to  the 4D Planck scale in M-theory units
\eq{
  \label{planckscale4D}
  M_{\rm pl}^2={\frac{1}{\pi}} M_*^9 R_{11} R_{\rm CY}^6
   ={\frac{1}{\pi}}M_*^2\, r_{11}\,
         r_{\rm CY}^6\sim M_*^2\, r_{11}\,,
}
where small letters indicate radii in units of $M_*$.
In the final step we have used the previous observation that
$r_{\rm CY}\sim 1.5$. Using this relation, we can now fix the 11D
Planck scale as
\eq{
             M_*\sim M_{\rm pl}^{2/3}\, R_{11}^{-1/3} \sim 10^9\, {\rm GeV}\,.
           }
Obviously, this is the same relation as \eqref{hannover96} and \eqref{werder} so that we
obtain the following scales
\eq{
  \label{thatisthelimit}
           M_*&\sim \Lambda_{\rm sp}\sim 10^9\,{\rm GeV}\,,\qquad
           M_s\sim M_{\rm pl}\sim 10^{19}\,{\rm GeV}\,,\qquad
           M_{\rm KK}\sim 10^{-12}\,{\rm GeV}\,,\\
           r_{\rm CY}&\sim 1.5\,,\qquad\phantom{aaiaaaaiiaa} r_{11}\sim 10^{20}\,.
         }
Note that except for the small value of $r_{\rm CY}$,  we are relaxed
here about precise numerical factors, as our emphasis is on parametric hierarchies.      
 
Before we proceed, let us  compare this to
a  GUT realization in HW theory \cite{Witten:1996mz}.
For the Dark Dimension Scenario,  we have solved the three relations
  \eq{
    \Lambda_{\rm DE}^{1/4}\sim {\frac{M_*}{ r_{11}}}\,,\qquad
    M_{\rm pl}^2\sim M_*^2 r_{11} r_{\rm CY}^6\,,\qquad
    {\frac{1}{\alpha_{\rm SM}}}\sim r_{\rm CY}^6
  }
  for the three unknowns $M_*, r_{11}, r_{\rm CY}$.
In contrast, in GUT realizations of HW theory \cite{Witten:1996mz} one solves the three relations
\eq{
        M_{\rm GUT}\sim {\frac{M_*}{ r_{\rm CY}}}\,,\qquad
    M_{\rm pl}^2\sim M_*^2 r_{11} r_{\rm CY}^6\,,\qquad
    {\frac{1}{\alpha_{\rm GUT}}}\sim r_{\rm CY}^6\,,
}
leading to  $M_{\rm GUT}\sim M_*\sim 10^{16}\,\rm{GeV}$
and a much smaller value $r_{11}\sim 10^{2-3}$. Including numerical
factors in \eqref{wittenrelation},  this was still at the boundary of $r_{11}^{\rm crit}>r_{11}$ being satisfied.
 
However, a critical distance
of order $r_{11}^{\rm crit} \sim 10^{2-3}$,
would spoil the Dark Dimension scenario in HW theory, unless we avoid warping by
choosing gauge bundles $V_1$ and $V_2$ satisfying
 symmetric tadpole cancellation 
\eq{
  \label{localtads}
   {\rm Tad_L}={\rm ch}_2(V_1)+{\frac{1}{2}} c_2(T_X)=
  -{\rm ch}_2(V_2)-{\frac{1}{2}} c_2(T_X)=0\,.
}

Looking at the heterotic gauge kinetic function \eqref{gaugekinf} for a non-abelian gauge
group
and expressing the scales of the various contributions in M-theory
units one schematically obtains
\eq{
  \label{gaugewithtad}
      4\pi  {\rm Re}f^{(i)}_{\rm YM}= r_{\rm CY}^6 \pm 4\pi^2\gamma (r_{\rm CY}^2 r_{11})\, {\rm Tad_L}
        +\mathcal{O}\big(e^{-2\pi r_{11} r_{\rm CY}^2}\big) +\mathcal{O}\big(e^{-2\pi r_{\rm CY}^6}\big) \,.
}
If we were not satisfying the tadpole condition on each $E_8$ wall,
the polynomial one-loop term    would be parametrically the  largest
one, not only leading to a strong
coupling singularity but also spoiling
the  previous relation \eqref{gaugescale4D}\footnote{The physical gauge
coupling is not equal to the real part of the gauge kinetic functions
but receives additional logarithmic corrections, see e.g.~\cite{Kaplunovsky:1995jw}, which
likely lead to  additional  $\mathcal{O}(1)-\mathcal{O}(10)$  contributions.
}.
We notice that for the values of
the radii  in the Dark Dimension Scenario \eqref{thatisthelimit}, the
$M2$- and $M5$-brane instanton
corrections in
\eqref{gaugewithtad} still seem to be controlled. This might be
interpreted as some evidence that this  regime is not
quantum obstructed.

In case of abelian fluxes,
one also has to require that the additional term in \eqref{gaugeu1}
for the hypercharge $U(1)$ is vanishing, as it would lead to a too small
value of the hypercharge gauge coupling.
In addition, after diagonalizing the $U(1)$ gauge coupling matrix,
all entries have to be  positive, i.e.
${\rm Re}f^{(i)}_{ab}$ must be  positive definite
with $\Delta{\rm Re}f^{(i)}_{YY}=0$.
Whether this is possible
depends on the global aspects of the concrete model and
needs to be studied case by case.

In view of the previous discussion of the gauge couplings, 
one may also examine the reliability of using the supergravity result \eqref{planckscale4D}  for the
4D Planck scale.
From the weakly coupled string perspective,
higher order curvature corrections could lead to corrections of the
4D Planck scale.
Let us schematically include such corrections in the 10D
action as
\eq{
      S\sim M_s^8 \int d^{10} x \sqrt{-G}\, {\frac{1}{g_s^2}} \left(  R +
        \sum_{n,l\ge 1} a_{l,n}
        {\dfrac{g_s^{2l}}{M_s^{2n}}}   R^{n+1} \right)\,\,,
    }
    where we included higher derivative and stringy  l-loop corrections 
    keeping track of orders of $M_s$ and $g_s$.
Upon dimensional reduction, these could contribute to the 4D Planck mass
\eq{\label{Mplcorr}
        M^2_{\rm pl} \sim M_*^2 r_{11} r_{\rm CY}^6 \left(1+\sum_{n,l\ge 1}
           a_{l,n} {\frac{r_{11}^{3l-n}}{r_{\rm CY}^{2n}} }\right)\,,
       }
where we took into account that the curvature scales like $R\sim
R_{\rm CY}^{-2}$.
It is now clear that a non-vanishing contribution with a positive power of
$r_{11}$, i.e.~for $3l>n$ would spoil the relation  \eqref{planckscale4D}.
Of course, it also matters whether the index structure
is such that one obtains the 4D Ricci scalar upon dimensional
reduction.

To our knowledge, such non-renormalization
theorems for higher curvature corrections in 10D
heterotic string theory have only been suggested for the $R^4$ term.
 Although for  $N=1$ supersymmetry it is not  1/2 BPS saturated, 
 it is indeed not being
 renormalized beyond
 one-loop \cite{Green:2016tfs} (see also the previous work \cite{Tseytlin:1995bi}).
Alternatively, one could directly consider corrections to the
 Planck scale in four dimensions. In this context, it has been shown
 in \cite{Florakis:2016aoi} that 
 the Planck mass does not  receive corrections  at one-loop level.
 In view of the complexity of calculating such higher curvature terms
 directly from string theory, 
 this analysis is not conclusive and other more direct methods are desirable.

\subsection{Comment on proton decay}

We have already mentioned that the Dark Dimension Scenario
does not seem straightforwardly compatible with a GUT scenario.  Let us
consider this issue in more detail.

Since the GUT scale is some orders of magnitude larger
than the species scale, it was argued in  \cite{Heckman:2024trz} that
the GUT gauge bosons should be  rather  solitonic strings of 5D Planckian tension
that extend through the dark dimension.
Moreover, it has been pointed out in \cite{Reig:2025dpz}  that there
are fundamental  tensions with experimental
bounds on proton decay challenging the combination of the
Dark Dimension  Scenario with HW theory.

The point is that in 10D one starts with $E_8$ and that the breaking
of this $E_8$ leaves traces in the massive spectrum.
Both breaking via discrete Wilson-lines $w$ and abelian gauge fluxes $F$
lead to massive vectors  in the bifundamental $(3,2)$ representation of
$SU(3)\times SU(2)$, whose mass scale is set by the KK scale
along the Calabi-Yau
\eq{
  \label{wilsonlandau}
            M^{\rm WL}_{(3,2)} \sim {\frac{w}{R_{(w)}}} \sim M_*{\frac{w}{
              r_{(w)}}}\,,\qquad\quad
            \left(M^{\rm flux}_{(3,2)}\right)^2 \sim \frac{F}{{\rm Vol}_{2,(F)}}
            \sim M_*^2\, \frac{F}{{\rm vol}_{2,(F)}}\,.
}
Here, $R_{(w)}$ denotes the radius of the 1-cycle
supporting the discrete Wilson line and ${\rm Vol}_{2,(F)}$ is
the volume of the 2-cycle supporting the quantized gauge flux $F$.
For super Planckian (in 5D) radii of the Calabi-Yau cycles,
the masses of these states are smaller than the 11D Planck scale  $M_*\sim 10^9\,$GeV and 
hence can generate too fast proton decay via dimension six
operators. One way to avoid this effect is that
the Calabi-Yau is not isotropic and contains cycles of sub-Planckian size.
 As the example of  a blow-up 2-cycle shows, this does not necessarily imply that
the volume is affected, but this regime might
be quantum obstructed and is definitely beyond any control.

For the case of flux breaking via line bundles there exists another
possibility, well known from the weakly coupled heterotic string.
Namely, that the line bundles lead to abelian gauge groups,
which generically are subject to the Green-Schwarz mechanism
by which some of them become massive and survive
as approximate global symmetries. Their breaking
occurs via instantons whose actions transform under
the gauged axionic shift symmetries. Hence one can exponentially
suppress dangerous proton decay operators (of any dimension $\Delta$)
by having e.g.~the sum of baryon and lepton number  $B+L$ as one of the
initial $U(1)$ gauge symmetries. Schematically, such operators take the form
\eq{
                        {\cal O}_{\slashed{B}}\sim\frac {e^{-
                            S_{\rm inst}}}{ M_*^{\Delta-4}}\,,
}
with $S_{\rm inst}=S_{N\!S5}=2\pi s$ for $N\!S5$-brane instantons
and  $S_{\rm inst}=S_{\rm F1}=2\pi \tau$ for world-sheet instantons.
Here we have defined $s={\rm Re}S=r_{\rm CY}^6$   and  $\tau={\rm
  Re}T=r_{11}r_{\rm CY}^2$.
Due to symmetric tadpole cancellation, the imaginary part of $S$
does not participate in the Green-Schwarz mechanism \cite{Blumenhagen:2005ga} and
therefore the proton decay operators are always suppressed
by world-sheet instantons 
which scale as $S_{\rm F1}=2\pi
r_{\rm CY}^2 r_{11}\sim \mathcal{O}(10^{20})$ in the HW limit. Hence,  we are on
the safe side regarding the rate of proton decay. Notice that without
tadpole cancellation on each $E_8$ wall, an $N\!S5$-brane instanton would possibly not provide
a sufficient suppression of dimension four proton decay operators.

In Appendix \ref{app_a}, based on \cite{Blumenhagen:2005ga} we provide a heterotic quiver for a
realization
of the MSSM spectrum with additional approximate global symmetries $B$
and $L$. This is based on a vector bundle background with structure group $S(U(3)\times
U(1)^3)\subset SU(6)$ and serves as a proof of principle
that such constructions are  a priori not obstructed.
Of course, to find a concrete Calabi-Yau threefold
and supported vector bundles satisfying all
the topological model building constraints is not a trivial task
and is beyond the scope of this work.

\subsection{The 5D M-theory limit}

Let us now focus on the regime
in parameter space where the relations \eqref{thatisthelimit} take us.
We infer that          
one is  driven to a parameter region very close
to an infinite distance point in the HW moduli space.
This limit can be characterized by the scalings
\eq{
       R_{11}\to \lambda R_{11}\,,\qquad R_{\rm CY}\to \lambda^{1/3}R_{\rm CY}\,,\qquad 
       M_*\to \lambda^{-1/3} M_*
}
for $\lambda\sim (M_{\rm pl}/\Lambda_{\rm DE})^{1/4}\sim10^{30}$.

\begin{figure}[ht]
    \centering
\includegraphics[width=0.85\linewidth]{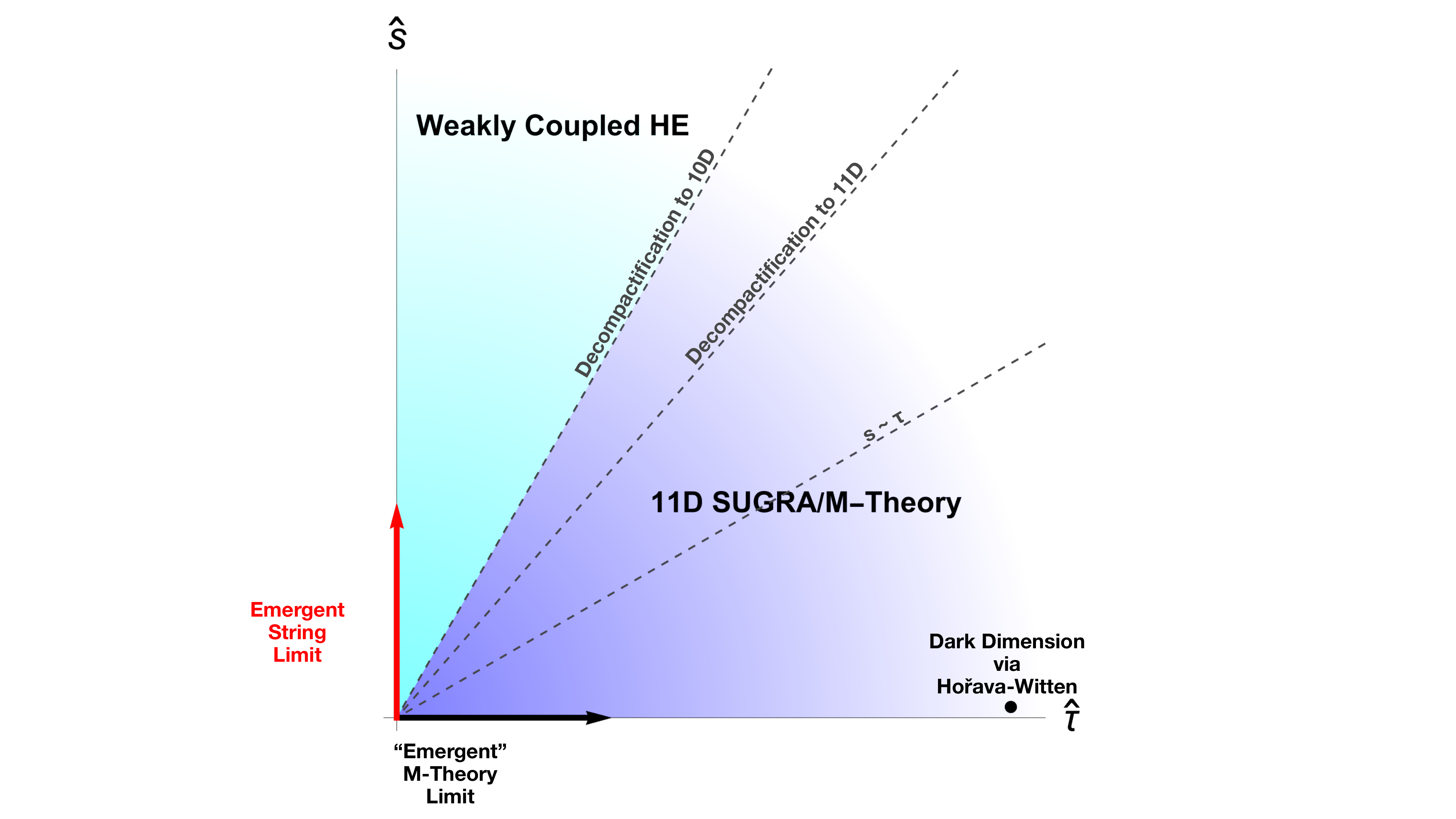}
    \caption{Moduli space of 4D $N=1$ heterotic $E_8\times E_8$ compactifications on an isotropic Calabi-Yau described by the canonically normalized scalars 
    corresponding to the universal saxion $\hat{s}$ and the (single) K\"ahler modulus $\hat{\tau}$.  
    The arrows indicate the heterotic emergent string limit (in red)
    and the 5D M-theory limit (in black). The point in moduli space
    corresponding to the HW realization of the Dark Dimension
    Scenario is indicated, as well as the decompactification
    directions to 10D and 11D. Along   $s\sim \tau$ we also show  the potential boundary of the classical moduli space when the eleventh direction would reach the critical size set by \eqref{wittenrelation}. For further details see also \cite{Grieco:2025bjy}.}
    \label{fig:4dN=1frames}
\end{figure}

To make the distinction between the various limits and duality frames
apparent, let us summarize them in Figure \ref{fig:4dN=1frames}. 
In \cite{Grieco:2025bjy}, the various perturbative infinite distance
limits of 4D $N=1$ theories were classified and related to different
EFT strings \cite{Lanza:2021udy,Lanza:2022zyg} becoming
light. Assuming an isotropic Calabi-Yau compactification, the various
limits can be described by the relative scalings of two moduli, the
universal saxion $s=\mathcal{V}_6/g_s^{2}$ and a single K\"ahler
saxion $\tau= \mathcal{V}_6^{1/3}$, where $\mathcal{V}_6$ is the
Calabi-Yau volume in string units.\footnote{In this setup the moduli space metric is flat and diagonal with the only non-zero components being \cite{Grieco:2025bjy}
\begin{equation*}
    G_{ss}=\frac{1}{2}\frac{1}{s^2}\,, \quad G_{\tau\tau}=\frac{3}{2}\frac{1}{\tau^2}\,,
\end{equation*}
leading to $s= e^{\sqrt{2}\hat{s}}$ and $\tau=e^{\sqrt{\frac{2}{3}}\hat{\tau}}$, where the hats indicate the canonically normalized counterparts.} 
This can be translated
to M-theory units so that 
$s= r_{\rm CY}^6$ and  $\tau=r_{11} r_{\rm CY}^2$.
Assuming that
$s,\tau>1$, we identify two different duality frames. The (parametrically) strong and weak coupling regimes
correspond to $s<\tau^3$ and $s>\tau^3$ respectively.

From figure \ref{fig:4dN=1frames} one can read-off four distinguished asymptotic limits $\lambda\to\infty$.
Along the vertical axis lies the perturbative
heterotic string limit, which in M-theory units is given
by the scalings
\eq{
       r_{11}\to \lambda^{-2/3} r_{11}\,,\qquad M_*\to  \lambda^{-2/3}
       M_*\,,\qquad   r_{\rm CY}\to \lambda^{1/3} r_{\rm CY}\,.
     }
Here the species scale is identical to the fundamental string scale $M_s$.
Along the horizontal axis  there is the aforementioned 5D
M-theory limit
\eq{
       r_{11}\to \lambda^{2/3} r_{11}\,,\qquad M_*\to  \lambda^{-1/3}
       M_*\,,\qquad   r_{\rm CY}\to \lambda^{0} r_{\rm CY}\,
     } 
with the species scale being the 5D Planck scale.     
From the figure it is apparent  that these two limits
are at the boundary of the two duality frames.

For completeness, we also mention the remaining two special
limits. 
Along the line $s=\tau^3$, there is the 10D
decompactification limit
\eq{
     r_{11}\to \lambda^{0} r_{11}\,,\qquad M_*\to  \lambda^{-1/2}
       M_*\,,\qquad   r_{\rm CY}\to \lambda^{1/6} r_{\rm CY}\,,
}
meaning that $g_s$ does not scale. One also has
the decompactification limit to 11D 
\eq{
     r_{11}\to \lambda^{2/9} r_{11}\,,\qquad M_*\to  \lambda^{-7/9}
       M_*\,,\qquad   r_{\rm CY}\to \lambda^{2/9} r_{\rm CY}\,.
}

The Dark Dimension Scenario via HW corresponds to the point
$(\hat s,\hat \tau)\sim (1.5,57)$ in  figure \ref{fig:4dN=1frames},
so that it is very close to the horizontal line and therefore to the boundary of validity of 11D supergravity.
We notice that this is the same asymptotic direction that
appeared in the recent
work \cite{Blumenhagen:2023tev,Blumenhagen:2023xmk,Blumenhagen:2024ydy,Artime:2025egu,Artime:2026kfq}
on the Emergence Proposal, 
which was a decompactification limit from  $D$ to $D+1$ dimensions.
In the following we call this limit the $(D+1)$-dimensional M-theory limit
to distinguish it from the complete eleven dimensional M-theory limit.

As shown in that series of papers, in this special limit
certain BPS protected couplings were completely  determined 
by integrating out the light towers
of states in a Schwinger integral.  From a weakly coupled string point of view,
all perturbative and non-perturbative
contributions in the string coupling were obtained.
Moreover, in analogy to perturbative string theory, the light towers of states 
were defined as those towers with a typical mass
scale not larger than the species scale, i.e.~the 11D
Planck scale.
Such a computation deep in the M-theory regime
was  only reliable due to the BPS nature of the couplings considered.
For reliably extending such considerations to HW theory and beyond supersymmetrically
protected quantities,
one would have to understand the quantization of M/HW theory.
While this is clearly beyond the scope of this work, we will point out some relevant conceptual key points. 

\section{Approaching the M-theory regime}
\label{sec:emergence}
 Despite the straightforward appearance of the desired hierarchies for a Dark Dimension realization  by HW theory, we are driven to
a regime in moduli space where neither perturbative string theory
nor 11D supergravity 
are truly trustable. In the absence of a quantized theory of M-theory, we
will content  ourselves with
approaching this problem from two certainly restricted perspectives,
first, as we also did in the previous section \ref{sec2}, the perturbative $E_8\times E_8$ heterotic string and second
the M-theoretic Emergence Proposal. In particular, we will
comment on contributions to the scalar potential, gauge couplings and Planck mass and how
they scale with respect to the radii $r_{11}$ and $r_{\rm CY}$.

\subsection{The scalar potential for the heterotic string }

Let us start the analysis of the scalar potential in the weakly coupled heterotic
string regime.  Since this is not a protected quantity we do not
expect that the functional behavior will be the same
 in the actually relevant 5D  M-theory limit. In the following, we
 will comment on the various differences in how certain quantities are
 expected to behave in the two limits. To make comparisons apparent,
 we will  translate expressions in the weakly coupled heterotic regime to M-theory units.

For the Dark Dimension scenario, the leading
order non-vanishing scalar potential (i.e.~the dark energy
in the quasi dS-vacuum) scales like \eqref{dSdistance}, i.e.
\eq{
                  V\sim {\frac{M_*^4}{r_{11}^4}} + \ldots\,.
}
Usually one argues that this comes from integrating
out  the tower of light KK modes at one-loop level.
However, in HW theory these are  KK modes along the
eleventh direction, i.e.~non-perturbative states
of mass $m\sim M_s/g_s$ in the heterotic string.
As such, they are not integrated out in an emergent
string limit at one-loop level. Instead, they are treated
as classical configurations leading to instanton corrections.
Hence, this contribution to the scalar potential
is not evident from the perturbative heterotic string.

Let us estimate the contribution of a world-sheet instanton
to the scalar potential in weakly coupled string theory.
This can be obtained from the 4D $N=1$ supergravity formula
for the F-term scalar potential
\eq{
  \label{Ftermpot}
           V_F\sim M_{\rm pl}^4\, e^K \left( G^{T\ov T} D_T W D_{\ov T}
               \ov W - 3|W|^2 \right)
}
with the K\"ahler potential
\eq{
           K=-\log (S+\ov S) - 3 \log (T+\ov T)\,.
}
We remind that the two moduli fields are the  4D dilaton    ${\rm Re}S=s=(M_s R_{\rm CY})^6/g_s^2$ and
the (single) K\"ahler modulus ${\rm Re}T=\tau=M^2_s R^2_{\rm CY}$.
A single world-sheet instanton (wrapping a rigid rational curve)
contributes a term $W\sim  A e^{- \pi T}$ to the superpotential
so that after changing to M-theory units we can estimate such a potential contribution at leading order as
\eq{
  \label{weakinstantons}
      V_F\sim M_*^4\, r_{11} r_{\rm CY}^4 \, e^{-2\pi r_{11} r^2_{\rm CY} }\,.
}
The exponent indicates that in HW theory, a world-sheet instanton
becomes a longitudinal $M2$-brane instanton, i.e.~it wraps
also the large eleventh direction.
Note that the prefactor contains the K\"ahler potential,
which is expected to receive string loop and $\alpha'$ corrections,
so that this cannot be trusted  as the leading order  result in the
5D M-theory limit. Some potentially relevant terms in the supergravity
limit of HW theory have been derived in \cite{Lukas:1998tt}.

\subsection{Emergence of M-theory limit}

As we have stressed so far, the relevant regime in the HW moduli space is
where $r_{\rm CY}\sim 1.5$ and $r_{11}$ exponentially large.
Thus, in order to study the physics and in particular (partial) moduli
stabilization for the Dark Dimension Scenario one needs
to compute the scalar potential in this
M-theory regime. We remind that for  $r_{\rm CY}=1.5$ this is not expected to  be safely captured
by the 11D supergravity approximation.

Nevertheless, in this peculiar infinite distance limit, it
has been established that the {\it exact} form
of certain 1/2 BPS saturated couplings in compactifications
of M-theory
are encoded in  Schwinger integrals
with all light towers of particle states running in the loop.
We remind that by light towers we mean all towers of 1/2 BPS states with
a typical mass scale not larger than the species scale.
In addition, one could also integrate out the heavy towers, but these
gave redundant contributions adding up to zero \cite{Blumenhagen:2024ydy,Artime:2025egu,Artime:2026kfq}.

In the HW case, one starts with at most $N=1$ supersymmetry
in 4D and therefore there are no BPS states.
However, we expect that
the species scale 
is still  the 5D respectively 11D
Planck mass $M_*$ so that from the former 1/2 BPS states
in M-theory, we expect to integrate over the remnants of  
all KK modes in the eleventh direction and all wrapped
transverse $M2$- and $M5$-branes \cite{Blumenhagen:2023xmk}.
Note that due to the large eleventh direction,
longitudinal branes are more massive than the species scale.
Since the full amplitude arises as a quantum effect
this behavior was called ``Emergence'' in the sense of
the Emergence Proposal.
In addition,  $M5$-branes wrapped over 4-cycles
in the Calabi-Yau also give rise to 4D strings with tension
of the order of $M^2_*$. For the considered 1/2 BPS
amplitudes these played no role in previous applications
which focused mainly on $d\geq 6$, but for non BPS amplitudes in 4D
they could be contributing,
though we certainly  do not know how to
quantize them.

In the following, we will assume that the M-theoretic Emergence Proposal
is correct and that this picture carries over
also to HW theory and non BPS saturated quantities, like
the scalar potential in  theories with
$N\leq1$ supersymmetry in 4D. With this in mind, we will speculate
about the resulting scaling properties of certain physical quantities.

\subsubsection*{Scalar potential}

Our ``Emergence" speculation means  that the (exact) vacuum energy in this regime
limit can be computed via the Schwinger integral
\eq{
       \rho= \frac{1}{2} M_*^4 \int_0^\infty {\frac{dt}{t}} \, {\rm
         Str}_{{\cal H}_L} \int {\frac{d^4k}{(2\pi)^4}}\,e^{-\pi t\, (k^2+m^2)}\,,
     }
where the supertrace is over the Hilbert space of the full light towers of states $\mathcal{H}_L$
and $m$ denotes their 4D masses in units of the eleven
dimensional Planck scale.\footnote{This formula is
  the same as for the one-loop
correction to the cosmological constant in perturbative
string theory, prior to implementing the level matching condition
for physical string states.
Here the light towers would just be KK, winding modes
and fundamental string excitations. Their typical
mass scale is equal to the string scale $M_s$, which is
the species scale in this infinite distance limit. In contrast to
the M-theory limit,
this would not give the full exact vacuum energy.}
Since we cannot quantize M/HW theory from first principles,
we neither know its Hilbert space nor its  spectrum.
At present, the best we can do is to estimate the contribution
from certain expected  states in ${\cal H}_L$.

Carrying out the 
Gaussian integral over the continuous 4D momenta $k_\mu$, yields
\eq{
       \rho\sim M_*^4 \int_0^\infty {\frac{dt}{t^3}} \, {\rm Str}_{{\cal H}_L} e^{-\pi t\, m^2}\,.
     }
Since the background is the direct product $X\times I_1$
and the KK momentum in $x_{11}$ is orthogonal to
the $M2$- and $M5$-branes,
it is reasonable that the mass formula has the form
\eq{
  \label{massdo}
       m^2=\left({\frac{n}{r_{11}}}\right)^2 + m_\perp^2\,,
}
where $n\in\mathbb Z$ denotes the KK momentum mode and $m_\perp$
the mass of the orthogonal excitations.
Due to the $\mathbb Z_2$ orbifold it is actually the invariant linear
combination of KK modes
\eq{
  \label{nonBPSKK}
  |n\rangle +  |-n\rangle
}
that is contributing. From the upstairs M-theory on $X\times S^1$
perspective, this is a non BPS configuration dual to an unstable
$D0$-$\ov{D0}$ brane.

Recently, in \cite{Kaufmann:2026tsy}   three reasons, called I,II, III,  were presented why such states
cannot be the light towers of states expected from the
swampland distance conjecture in certain infinite distance limits of  4D $N=1$ supersymmetric type IIA
  orientifold models. 
  We expect that in our  5D M-theory limit they do not
  apply\footnote{We thank L. Kaufmann, T. Weigand and M. Wiesner
    for enlightening discussions on this point.},
  as we have a trivial fibration (III) and, since the $D0$-$\ov{D0}$ states
  geometrize like \eqref{nonBPSKK} in the strong coupling limit,
  these states are expected to be asymptotically stable (II)
  and that their mass in that limit is really given by the first term
  in \eqref{massdo} (I).

Now, isolating  the contribution to the vacuum energy
where $m_\perp=0$, we can regularize the UV divergence at small $t$
with the methods from \cite{Blumenhagen:2024ydy} and obtain
\eq{
  \label{SchwingerKK}
       \rho_{\rm KK}\sim M_*^4 \int_0^\infty {\frac{dt}{t^3}} \,
       \sum_{n\ne 0} e^{-\pi t {\frac{n^2}{ r_{11}^2}}}\sim
       {\frac{M_*^4}{r_{11}^4}}\sim  {\frac{M_{\rm pl}^4}{\tau^6}}\,
     }
     with   the K\"ahler modulus $\tau=r_{11} r_{\rm CY}^2\gg 1$.
This is indeed the expected contribution to the vacuum energy
from integrating out the lightest tower of states.
In contrast to the   weakly coupled heterotic string,
here the KK modes in $x_{11}$ are considered as perturbative
states and are to be integrated out.
Let us emphasize that in this and the upcoming  computations we  ignore
potential cancellations between bosonic and fermionic
contributions.

To get an impression of what one might obtain in addition,
let us consider again an isotropic Calabi-Yau and therefore  a single K\"ahler modulus.
Then, one can consider a  transverse $M2$-brane wrapping a 2-cycle.
More concretely, since the orbifold action includes $C_3\to -C_3$,
from the upstairs perspective, one gets a linear combination of a (bulk) $M2$-brane and an
image anti $M2$-brane. Indeed, the invariant combination would be
\eq{
  \label{nonBPSM2}
           |M2, x_{11}, p_{11}\rangle + |\ov{M2}, -x_{11}, -p_{11}\rangle\,,
}
which might develop  a tachyon for $x_{11}$  close to the $\mathbb Z_2$ fixed
points.  However, in the bulk of $x_{11}$ it gives a stable non-BPS state
which we expect  to be present in the HW Hilbert space. 
As a word of caution, since these are non-supersymmetric objects in
M/HW-theory, one cannot expect them
to be  correctly counted by just the homology group.

Boldly assuming that, maybe up to some strong coupling effects  in the numerical
coefficients,  the mass formula still takes the simple BPS-like form
\eq{
  \label{KKandM2}
       m^2=\left({\frac{n}{r_{11}}}\right)^2 + \left({p\, r^2_{\rm
             CY}}\right)^2+  m_\perp^2\,,
}
for $m_\perp=0$ this would lead to a contribution (the $p=0$  case is given by \eqref{SchwingerKK})
\eq{
       \rho_{{\rm KK}+M2}\sim M_*^4 \int_0^\infty {\frac{dt}{t^3}} \,
       \sum_{\substack{p\neq 0\\{n\in \mathbb{Z}}}}  \alpha_p\, e^{-\pi t \left({\frac{n^2}{r_{11}^2}} +
          (p \, r_{\rm CY}^2)^2 \right)}\,.
     }     
Here, $\alpha_p$ counts the number of contributing 2-cycles and we
assumed  that the 2-cycle is in the bulk of
$x_{11}$ so that one can ignore contributions to their masses from
the gauge flux on the end-of-the-world walls.
This expression is reminiscent of the Gopakumar-Vafa
computation \cite{Gopakumar:1998ii,Gopakumar:1998jq} for
corrections to the prepotential in $N=2$ compactifications in 4D and 
can be evaluated straightforwardly\footnote{First, one performs a Poisson resummation in $n$ and uses 
the integral representation \cite{Kiritsis:1997em}
\eq{
\label{besselrel}
\int_0^\infty \frac{dx}{x^{1-\nu}} \,e^{-{\frac{b}{x}}-cx}=2 \left|  {\frac{b}{c}}\right|^{\frac{\nu}{2}} K_\nu\left(2\sqrt{|b\, c|}\right)\, ,\nonumber
}
where $K_\nu(x)$ denotes the modified Bessel-function of order $\nu$.
Moreover, we have employed  the finite expansion $ K_{\pm {\frac{5}{2}}}(x)=\sqrt{\frac{\pi}{2}}
  {\frac{e^{-x}}{\sqrt{x}}} \left(1+\frac{3}{x} +\frac{3}{x^2}\right)$. }
\eq{
  \label{rhofinale}
  \rho_{{\rm KK}+M2}\sim \beta\, {\frac{M_{\rm pl}^4}{\tau}}   +
   \frac{M_{\rm pl}^4}{\tau^4}    e^{-2\pi  \tau}\left(1+ \frac{3}{2\pi  \tau}
       +\frac{3}{(2\pi  \tau)^2}\right) + \mathcal{O}\big(e^{-4\pi  \tau}\big)
}
with $\beta$ a to be regularized  numerical coefficient and where we expressed
the result in terms of the 4D Planck scale. Remarkably, 
in this way
the potential only depends on the K\"ahler modulus $\tau$, which
is also   the action of an Euclidean
$M2$-brane wrapping the segment $I_1$ and a 2-cycle in the Calabi-Yau.

Even though this result was certainly derived in a too simplified
manner, it shows a couple of notable points.
Importantly, it confirms that such Schwinger integrals can
capture non-perturbative, i.e.~exponential,  contributions.
More concretely, one can compare the single instanton contribution
in \eqref{rhofinale} with the one
arising from a (probably too) simple supergravity set-up
with one complex modulus $T$ with ${\rm Re}T=\tau$ and  the K\"ahler-  and superpotential
\eq{
      K=-6\log(T+\ov T) \,,\qquad\qquad  W=A\, e^{-\pi  \, T} \,.
}
This leads  to the F-term potential 
\eq{
  \label{ftermdings}
   V_F\sim |A|^2   \frac{M_{\rm pl}^4}{\tau^4} \, e^{-2\pi  \tau}\left(1+ \frac{12}{2\pi  \tau}
       +\frac{18}{(2\pi  \tau)^2}\right)   \,,
}
where we have evaluated \eqref{Ftermpot} for the single modulus $T$.
Apparently, the similarity in structure is striking,
it is only the numerical factors that do not perfectly match.

Comparing the prefactor of the leading order term to the weak coupling result
\eqref{weakinstantons} we find a different  power of $r_{11}$, indicating that one
cannot simply extrapolate the weak coupling formula
to the 5D M-theory limit. In fact, relative to the latter,
the result \eqref{rhofinale} suggests a larger suppression at large values of
$r_{11}$.
Note that the first  term in \eqref{rhofinale} dominates over \eqref{SchwingerKK} and
gives the leading  contribution $\rho^0_{\rm bulk}\sim M_*^4 r_{11}
r_{\rm CY}^{10}\sim M_{\rm pl}^4/(r_{11} r_{\rm CY}^{2})$. 
Clearly, since in the Dark Dimension Scenario one wants that
the vacuum energy is given by
\eqref{SchwingerKK}, all such more dominant contributions to $\rho$
must cancel in the course of (partial) moduli stabilization.
Note that if such terms were not  there, one would have solved
the cosmological constant problem in this setup!

\subsubsection*{Gauge couplings and 4D Planck scale}

Finally, we speculate  on the emergence of the 4D gauge coupling and
the 4D Planck scale. 
Concerning the first, we would like to see whether one can get
the functional dependence of the four types of terms appearing in the
gauge kinetic function \eqref{gaugekinf}.
As for the 4D one-loop threshold corrections in string theory \cite{Kaplunovsky:1987rp},
we propose the following schematic form of  the Schwinger integral 
\eq{
  \label{Schwingergauge}
       {\rm Re} f_a \sim  \int_0^\infty {\frac{dt}{t}} \, {\rm Str}_{{\cal H}}
       \left(Q_a^2 \, e^{-\pi t\, m^2}\right)\,\,,
     }
where, as indicated by $Q_a^2$,  the trace is expected to only involve the towers of states  that
are charged under the corresponding gauge group.
Since the gauge fields live on the two $E_8$ walls the transverse
$M2$-$\ov{M2}$ branes in the bulk are neutral and
cannot contribute to the trace.  Instead the only charged extended objects we know
are strings, i.e.~longitudinal membranes which can give particle-like
states in 4D upon wrapping a 1-cycle of the CY threefold\footnote{These states have masses
$M\sim M_* r_{11} r_{\rm CY}$ and are heavier than the species scale
$M_*$. This suggests that for boundary contributions, we cannot
restrict the Schwinger integral to the light towers.}. However, smooth CY threefolds
do not contain any homological 1-cycles so one might wonder
how such states could potentially appear. 
We expect that either torsional
1-cycles contribute or that  membranes/strings  wrapping  exact 1-chains,
i.e.~$\Gamma_1=\partial \Gamma_2$,
are  stabilized by the gauge flux on the boundary walls. 
In addition, we had the  towers of Wilson line and Landau level states from
\eqref{wilsonlandau} which are also heavy charged excitations.
All these states can be thought of as $\mathbb Z_2$ twisted sector
states that cannot move away from the end-of-the-world branes,
meaning that they cannot carry 11D momentum.

The Schwinger  integral can now be carried out analogously to the one for the
scalar potential yielding
\eq{
       {\rm Re}f_{a,(1)}\sim \int_0^\infty {\frac{dt}{t}} \,
       \sum_{n\in \mathbb Z} e^{-\pi t \left({\frac{n^2}{r_{\rm CY}^2}} +
           (r_{11} r_{\rm CY})^2 \right)}\sim  \tau + {\cal O}\big(e^{-2\pi \tau}\big)\,,
     }
 with $\tau=r_{11} r_{\rm CY}^2$.     
Apparently, this yields the two terms from the stringy one-loop threshold
correction.  For the heterotic string,
such one-loop thresholds have been computed e.g.~for toroidal
orbifolds \cite{Dixon:1990pc,Mayr:1993mq}
and indeed they show the above behavior.  In fact, they arise from the
KK and winding modes in  so-called $N=2$ supersymmetric sectors of the orbifold.

In order to obtain the two $s$-dependent terms in eq.\eqref{gaugekinf} in a similar manner,
one only needs to assume that, similar to the 1-chains on the walls,
there also exist dual 5-chains along which a wrapped $M5$-brane
gets stabilized. This would lead to a Schwinger integral
\eq{
       {\rm Re}f_{a,(2)}\sim \int_0^\infty {\frac{dt}{t}} \,
       \sum_{n\in \mathbb Z} e^{-\pi t \left({\frac{n^2}{r_{\rm CY}^2}} +
           (r_{\rm CY}^5)^2 \right)}\sim  s+ {\cal O}\big(e^{-2\pi s}\big)\,
     }    
with $s=r_{\rm CY}^6$.  Recall that in the 5D M-theory limit, the
leading order term would be the linear term in $\tau$, whose
coefficient is actually vanishing by symmetric tadpole cancellation.
This was a prerequisite for being able to extend the moduli space
beyond the $s \sim \tau$ boundary (see also Figure \ref{fig:4dN=1frames}).

As a further speculation, let us consider the 4D Planck scale and revisit the
relation \eqref{planckscale4D} from the point of view of emergence of the
5D M-theory. 
Interpolating the measure factors in the Schwinger integrals for the $R^4$ and
the scalar potential ($\mathcal{O}(R^0)$), we would like to propose that
the coefficient of the Einstein-Hilbert term is given
by the Schwinger integral\footnote{In analogy to the gauge coupling
  \eqref{Schwingergauge},  this can also be written as  ${\cal M}^2
  \sim \int_0^\infty\! {\frac{dt}{t}}\,  {\rm Str}_{{\cal H}}
  \big((M_*m)^2e^{-\pi t m^2}\big)$.}
\eq{
  \label{Schwingermpl}
       {\cal M}^2 \sim M_*^2 \int_0^\infty {\frac{dt}{t^2}} \, {\rm
         Str}_{{\cal H}}\, e^{-\pi t\, m^2}\,.
     }
First we consider the contribution of the 11D KK modes \eqref{nonBPSKK} and the  bulk wrapped
$M2$-$\ov{M2}$-branes \eqref{nonBPSM2}, giving 
the leading order contribution (for large $r_{11}$)
\eq{
         {\cal M}_{(0)}^2 \sim M_*^2\, r_{11}\, r_{\rm CY}^6  \sim
         M_{\rm pl}^2\, ,
 }      
which is precisely the wanted expression for the 4D Planck scale
$M_{\rm pl}^2$!
However, we expect  also the previous two pairs of charged states on the $E_8$ walls
to contribute to the Schwinger integral.
Their leading order contributions are
\eq{
      {\cal M}_{(1)}^2 \sim M_*^2\, r_{11}^3\, r_{\rm CY}^4  \sim
     {M_{\rm pl}^2}\frac{\tau^2}{s}\,,\qquad
      {\cal M}_{(2)}^2 \sim M_*^2\, r_{\rm CY}^{16}  \sim
         M_{\rm pl}^2\frac{s^2}{\tau}\,.
       }
 Clearly, in the 5D M-theory limit,    ${\cal M}_{(1)}$ is
 parametrically  larger than the wanted Planck scale
 and for a negative sign  could also signal  an obstruction(!)
 for this limit.
 Hence, the consistency of the whole construction depends
 on whether the coefficient in front of it is vanishing.
 Whether this is the case, we don't know but since
 the similar effect happened for  the gauge coupling due to
 symmetric tadpole cancellation, it
 might also happen here.

We reiterate that these considerations are touching upon
yet to be understood regions of the HW moduli space. Our certainly incomplete results 
indicate that one may still be optimistic about obtaining at least the scalings of various
couplings from Schwinger integrals. At the same time, we also highlighted some of the open problems related to our approach.

\section{Conclusions}

We have collected  more evidence  that the realization of the
Dark Dimension Scenario via Ho\v{r}ava-Witten (HW) theory might be a very
attractive and natural possibility. We argued that some of the
concerns raised about such a construction, like the too strong backreaction of
the $E_8$ walls or  too fast proton decay, can potentially be circumvented
by allowing
more general background gauge fluxes including line bundles
and by implementing symmetric tadpole cancellation on the two $E_8$ walls.
This is a natural condition, as then the warp factor does not depend on
the eleventh coordinate  and one   keeps the wanted 
product structure $X\times I_1$ intact.
Whether this more tuned construction is indeed a loop-hole for 
the appearance of quantum obstructions \cite{Kaufmann:2026fli,Kaufmann:2026mha,Kaufmann:2026tsy}
in the large $r_{11}$ limit  of HW theory, remains to be seen.
Concerning proton decay, as a proof of principle, in the appendix we present a heterotic quiver 
realizing  the MSSM spectrum extended by two approximate
global symmetries corresponding to the lepton and baryon number.
It would be interesting to explore whether full fledged  heterotic/HW
models of this type can indeed be constructed.

As a generic feature of such a realization
of the Dark Dimension Scenario, we emphasized that
one is driven to a specific large distance  corner  of HW theory,
where the M-theoretic Emergence Proposal had been shown to apply
at least in the  simpler context of 1/2 BPS saturated couplings.
A speculative  extrapolation of  these results
suggested that the exact vacuum energy/scalar potential
as well as  the gauge couplings and the 4D Planck scale
can be given by   Schwinger integrals over the
(light) towers of states.  While their precise
evaluation is certainly not possible without understanding
the quantization of M-theory and the resulting
spectrum, we have made a few steps towards extracting at least
partial information from it.

To summarize the main message of this work:
The realization of the Dark Dimension Scenario in HW theory seems
to be a very natural possibility, but forces us
to understand quantum gravity in a specific regime of strong coupling. It therefore presents the intriguing
possibility that a realization of the Dark Dimension scenario is
closely tied to understanding
the quantization of M-theory. Although this is a formidable task, our goal was to highlight that the extrapolation of the
Emergence Proposal can be useful in obtaining some preliminary results.

\paragraph{Acknowledgments.}
We would like to thank Alessandra Grieco, Lukas Kaufmann, Andriana Makridou,
Stephan Stieberger,  Timo Weigand  and  Max Wiesner  for useful
discussions and Manuel Artime and Niccol\`o
Cribiori for valuable comments on the manuscript.

\vspace{0.5cm}
\appendix

\section{Heterotic quiver with proton stability}
\label{app_a}

In this appendix, we present an example of a heterotic
quiver model that contains the MSSM spectrum and has
two additional approximate global $U(1)$ symmetries
that can be identified with lepton and baryon number.
Therefore, there are no perturbative operators
inducing proton decay.  Since for symmetric tadpole
cancellation, the global symmetries
are always  broken by world-sheet instantons,
the $B+L$ violating operators are exponentially suppressed
and can be consistent with experimental bounds.

This quiver model can be considered as  a  heterotic analogue
of the ``Madrid Quiver'' \cite{Cremades:2002te}, which provided the charge embedding
of the Standard Model into intersecting D6-branes
in type IIA orientifolds, i.e.~we are not specifying a concrete
Calabi-Yau threefold supporting vector bundles satisfying
all the topological conditions needed to realize such
a quiver. But the quiver shows that it is in principle
possible to embed the MSSM with lepton and baryon
number symmetries into one $E_8$ factor.

The quiver is constructed from one of the models
presented in \cite{Blumenhagen:2005ga}, in particular from the class
of models starting with an $S(U(3)\times U(1)^3)\subset SU(6)$ bundle
leading to the commutant gauge group $SU(3)\times SU(2)\times U(1)^3$.
The Table 7  and Table  10 from that paper contained
a list of 
potential MSSM particle embeddings and linear combinations
of the three initial abelian gauge factors $U(1)_Z$, $U(1)_X$,
$U(1)_{Y'}$
which could be identified with the hypercharge $U(1)_Y$.
In general all three $U(1)$'s receive  a mass via Green-Schwarz terms
so that one has to implement one condition on the vector bundles
to keep $U(1)_Y$ massless.  All these details are described
in \cite{Blumenhagen:2005ga} so that here we can just  focus on the data of the
quiver. 

We take a model from the first column in Table 7 for which the
hypercharge is
\eq{
    {\rm hypercharge}\quad    U(1)_Y&={\frac{1}{2}} U(1)_Z +{\frac{1}{10}} U(1)_X -\frac{1}{15}
    U(1)_{Y'}\,.
  }
  Then, inspection reveals that the realization of the MSSM spectrum
  as shown in Table \ref{MSSMdata} below has nice features.      
\begin{table}[ht]
    \begin{center}
    \begin{tabular}{|c|c| c| c c  c |c|} 
    \hline
     Fields & label  & rep & $U(1)_Y$ & $U(1)_L$ & $U(1)_B$ & vector bundle \\ 
    \hline
    $Q$ & D4 & $(3,2)_{(1,-1,1)}$ & $1/3$  & $0$ & $1/3$ & $E=V$ \\
    $U$ & C4 & $(\ov 3,1)_{(-2,-2,2)}$ & $-4/3$  & $0$ & $-1/3$ &
                                                                  $E=V\otimes L_1^{-1}$ \\
      $D$ & C1 & $(\ov 3,1)_{(0,4,-4)}$ & $2/3$  & $0$ & $-1/3$
                                             & $E=L_2^{-1}\otimes L_3^{-1}$ \\
    $L$ & B3 & $(1,2)_{(-2,2,3)}$ & $-1$  & $1$ & $0$ &
                                              $E=V\otimes L_1^{-1}\otimes L_2^{-1}$ \\
    $E$ & $\ov{\rm A3}$ & $(1,1)_{(3,1,-6)}$ & $2$  & $-1$ & $0$ &
                                                     $E=L_1\otimes L_3^{-1}$
      \\
      $N$ & ${\rm A4}$ & $(1,1)_{(1,-5,0)}$ & $0$  & $-1$ & $0$ &
                                                       $E=V\otimes L_2$
      \\
      \hline
    $H_u$ & B4 & $(1,2)_{(1,3,-3)}$ & $1$  & $0$ & $0$ & $E=V\otimes L_2^{-1}\otimes L_3^{-1}$ \\
    $H_d$ & $\ov{\rm B4}$ & $(1,2)_{(-1,-3,3)}$ & $-1$  & $0$ & $0$ & $E=V\otimes L_2^{-1}\otimes L_3^{-1}$ \\     
    \hline
    \end{tabular}
    \caption{Standard Model quiver  of the $E_8\times E_8$ heterotic string with
      an $S(U(3)\times U(1)^3)$ bundle and a commutant  initial gauge
      group $SU(3)\times SU(2)\times U(1)^3$.}
    \label{MSSMdata}
    \end{center}
  \end{table}
  There, the second column denotes the origin of the massless
  left-handed states as shown in Table 10 of \cite{Blumenhagen:2005ga}. The third column
  shows the $SU(3)\times SU(2)\times U(1)_Z\times U(1)_X\times
  U(1)_{Y'}$ charges and the  last column the associated vector
  bundle, whose Euler characteristic
  \eq{
    \label{euler}
    \chi(X,E)=\int_X {\rm ch}_3(E) +{\frac{1}{12}} c_1(E)\wedge c_2(T_X)
  }
  determines the chiral spectrum.
  From the charges in Table  \ref{MSSMdata} one directly infers that all four Yukawa couplings are charge neutral 
  \eq{
         Q\, U\, H_u\,,\qquad Q\, D\, H_d \,,\qquad L\, E\, H_d \,,\qquad L\, N\, H_u 
       }
  and are therefore perturbatively allowed.
 Moreover, besides the hypercharge one can define the two
linear combinations
\eq{
      {\rm lepton\ number}\quad  U(1)_L&={\frac{1}{5}} U(1)_X +{\frac{1}{5}} U(1)_{Y'}\\
        {\rm baryon\ number}\quad  U(1)_B&={\frac{1}{4}} U(1)_Z +{\frac{1}{20}} U(1)_X +{\frac{2}{15}}
         U(1)_{Y'}
       }
 with respect to which the MSSM particles in Table \ref{MSSMdata} carry lepton and baryon
 number charges. Together with the hypercharge, these are shown in column 4,5,6.

 They are assumed to gain a mass via appropriate Green-Schwarz
 couplings  and survive as perturbative global symmetries.
 As a direct consequence, all dimension $\Delta=4,5,6$ proton decay
operators are not charge neutral.
This is shown in more detail in Table \ref{protondecay}.
 \begin{table}[ht]
    \begin{center}
    \begin{tabular}{|c c| c| c c  c|  } 
    \hline
     Dim & ${\cal O}_{\slash\!\!\!\! B}$  & $U(1)_Z\times U(1)_X\times
  U(1)_{Y'}$  & $U(1)_Y$ & $U(1)_{B-L}$ & $U(1)_{B+L}$   \\ 
    \hline
    $\Delta=4$ & $QDL$ & ${(-1,5,0)}$ & $0$ & $-1$  & $1$  \\
       & $UDD$ & ${(-2,6,-6)}$ & $0$ & $-1$  & $-1$  \\
      \hline
      $\Delta=5$ & $QQQL$ & ${(1,-1,6)}$ & $0$ & $0$  & $2$  \\
         & $UUDE$ & ${(-1,1,-6)}$ & $0$ & $0$  & $-2$  \\
       & $UDDN$ & ${(-1,1,-6)}$ & $0$ & $0$  & $-2$  \\
      \hline
       $\Delta=6$ & $Q\ov{D}\, \ov{U} L$ & ${(1,-1,6)}$ & $0$ &
                                                                       $0$  & $2$  \\
         & $U\ov{Q}\,\ov{Q} E$ & ${(-1,1,-6)}$ & $0$ & $0$  & $-2$  \\
       & $D\ov{Q}\,\ov{Q} N$ & ${(-1,1,-6)}$ & $0$ & $0$  & $-2$  \\
      \hline
    \end{tabular}
    \caption{Proton decay operators and their $U(1)$ charges.}
    \label{protondecay}
    \end{center}
  \end{table}

  Ideally, for the charged matter fields one should get
  three generations, i.e.~$\chi(X,E)=3$,  accompanied by a single vector-like
  Higgs field with  $\chi(X,E)=0$.  The right handed neutrino $N$
  could also be absent, in which case $U(1)_{B-L}$ becomes anomalous.
  From Table 10 of \cite{Blumenhagen:2005ga} one infers
  that there are more potentially massless states, even chiral ones.
  Following the same logic as in \cite{Anderson:2014hia}  we have checked that
  implementing the MSSM spectrum shown in Table \ref{MSSMdata}
  is consistent with the expressions for the Euler characteristics \eqref{euler}
of the associated bundles, however inevitably leading
to extra chiral exotics. The appearance of the latter
is a genuine problem of string model building using
$U(1)$ bundles or intersecting D-branes, respectively.

Finally, for a complete construction of a bona fide MSSM model in the
Dark Dimension context, one would also need to
provide a suitable  Calabi-Yau manifold and two vector bundles
embedded into the two $E_8$ factors satisfying
symmetric tadpole cancellation.

\newpage

\bibliography{references} 
\bibliographystyle{utphys}

\end{document}